\documentclass[conference]{IEEEtran}
\IEEEoverridecommandlockouts
\usepackage{cite}
\usepackage{amsmath,amssymb,amsfonts}
\usepackage{algorithmic}
\usepackage{graphicx}
\usepackage{textcomp}
\usepackage{xcolor}
\usepackage{subfig}
\def\BibTeX{{\rm B\kern-.05em{\sc i\kern-.025em b}\kern-.08em
    T\kern-.1667em\lower.7ex\hbox{E}\kern-.125emX}}
\begin{document}

\title{TrollHunter2020: Real-Time Detection of Trolling Narratives on Twitter During the 2020 US Elections}

\makeatletter
\newcommand{\linebreakand}{%
  \end{@IEEEauthorhalign}
  \hfill\mbox{}\par
  \mbox{}\hfill\begin{@IEEEauthorhalign}
}
\makeatother

\author{\IEEEauthorblockN{Peter Jachim}
\IEEEauthorblockA{\textit{College of Computing and Digital Media} \\
\textit{DePaul University}\\
Chicago, IL, US \\
pjachim@depaul.edu}
\and
\IEEEauthorblockN{Filipo Sharevski}
\IEEEauthorblockA{\textit{College of Computing and Digital Media} \\
\textit{DePaul University}\\
Chicago, IL, US \\
fsahrevs@depaul.edu}
\linebreakand
\IEEEauthorblockN{Emma Pieroni}
\IEEEauthorblockA{\textit{College of Computing and Digital Media} \\
\textit{DePaul University}\\
Chicago, IL, US \\
epieroni@depaul.edu}
}

\maketitle

\begin{abstract}
This paper presents \textit{TrollHunter2020}, a real-time detection mechanism we used to hunt for trolling narratives on Twitter during the 2020 U.S. elections. Trolling narratives form on Twitter as alternative explanations of polarizing events like the 2020 U.S. elections with the goal to conduct information operations or provoke emotional response. Detecting trolling narratives thus is an imperative step to preserve constructive discourse on Twitter and remove an influx of misinformation. Using existing techniques, this takes time and a wealth of data, which, in a rapidly changing election cycle with high stakes, might not be available. To overcome this limitation, we developed \textit{TrollHunter2020} to hunt for trolls in real-time with several dozens of trending Twitter topics and hashtags corresponding to the candidates' debates, the election night, and the election aftermath. \textit{TrollHunter2020} collects trending data and utilizes a correspondence analysis to detect meaningful relationships between the top nouns and verbs used in constructing trolling narratives while they emerge on Twitter. Our results suggest that the \textit{TrollHunter2020} indeed captures the emerging trolling narratives in a very early stage of an unfolding polarizing event. We discuss the utility of \textit{TrollHunter2020} for early detection of information operations or trolling and the implications of its use in supporting a constrictive discourse on the platform around polarizing topics.
\end{abstract}


\section{Introduction}
The political disruptions in the U.K. and U.S. in 2016 brought widespread attention to the dissemination of false information online \cite{Sanger, Sirivianos}. The false or otherwise unverified information enables  ``alternative narratives'' to proliferate on social media about polarizing topics like elections, man-made disasters, or pandemics \cite{Starbird}. For example, the ``fake news'' that ``Pope Francis endorsed Donald Trump for president'' runs against the mainstream narrative of Pope Francis's message against Trump's politics of fear during his 2016 campaign \cite{Yardley}. The ``conspiracy theory'' that the ``Boston Marathon Bombings were perpetrated by U.S. Navy Seals'' counters the mainstream narrative of a terrorist attack perpetrated by foreign nationals \cite{Parker}. Or the ``rumor'' that the COVID-19 virus originated in a laboratory in Wuhan contradicts the mainstream narrative that the virus originated in nature ~\cite{Sanger}. 

Due to a lack of editorial judgement, fact-checking, or third-party filtering, social media in general, and Twitter in particular, are especially conducive to dissemination of alternative narratives ~\cite{Allcott, Caulfield}. With the possibility of re-tweeting and linking content with hashtags, Twitter also allows for a special case of alternative narratives as part of information operations campaigns with the goal to provoke emotional response from individual users when discussing polarizing topics \cite{Coles}. We refer to these alternative narratives as \textit{trolling narratives}. Trolling narratives incorporate fake news, conspiracy theories, and rumors \cite{Caulfield}, but also personal opinions, comments, memes, and provoking hashtags \cite{Broniatowski, Stewart}. 

In response to the dissemination of false information throughout 2016 \cite{Benkler, Cresci}, Twitter began to publish information operations datasets containing trolling narratives from various state-sponsored troll farms \cite{Twitter}. Since then, the detection of political information operations developed into a serious problem because the state-sponsored troll farms persisted in their efforts, using both fake and bot accounts, and used a wide array of polarized topics to choose from in creating trolling narratives. Usually, moderators review suspected trolling activity to ban/mute trolling users and flagging/deleting trolling content, but this kind of manual solution has some major drawbacks, including a delay of actions, subjectivity of judgment, and scalability \cite{Ortega, Fornacciari}. The need for automated trolling detection on Twitter thus drew the attention of the research community yielding various detection approaches \cite{Jane, Zannettou, Llewellyn, Ghanem}. 

Most of the existing automated trolling detection approaches utilize the information operations datasets from Twitter \cite{Twitter}. Although these approaches give valuable insight, they do not take into account the evolving nature of the trolling narratives with the development and introduction of polarizing topics (e.g. COVID-19 trolling \cite{nspw2020}) nor do they consider the readjustment and pivoting of trolling tactics in response to the widespread attention to social media information operations after 2016 (e.g. the introduction of Twitter misinformation labels \cite{Roth}). State-sponsored trolls, aware that there is an active hunt for dissemination of false information on Twitter, will most likely avoid using the trolling narratives and tactics from 2016 \cite{Spring}. Moreover, given the short time span between posting trolling content and being detected/flagged/removed by Twitter, the trolls will utilize trending topics to capture the moment and muddy discourse in real time (e.g. spreading rumors that Scranton, PA is not the real birthplace of Joe Biden during the vice-presidential campaign in 2020). 

To address these discrepancies, we developed a system for \textit{real-time troll hunting} that captures the contextual development of trolling narratives as they are disseminated on Twitter. Our system, called \textit{TrollHunter2020}, leverages a novel application of a correspondence analysis to hunt trolling narratives in real time during the 2020 U.S. election cycle, throughout the campaigning period as well as the week of the election. This novel approach provides decision support to an analyst using exploratory techniques to uncover hidden patterns in data. Our contribution of a mechanism for early detection of trolling narratives is a direct response to the imperative for increasing understanding of how alternative narratives evolve within the context of a polarizing topic such as the 2020 U.S. elections. Because\textit{TrollHunter2020} is intended for use in real-time as an arbiter of developing and intense discourses on Twitter, it raises important ethical concerns about its potential (mis)use and we included an ethical treatise of doing real-time trolling detection research.

\section {Automated Detection of Trolling Narratives}
\subsection{2016 U.S. Elections}
The automated trolling detection runs, to a certain extent, counter to social media platforms' goal to allow for a high degree of desired participation and constructive public discourse, making them reluctant to immediately exclude users exhibiting trolling behaviour to avoid perceptions of excessive control and censorship \cite{Fornacciari}. However, the need for automated detection of trolling narratives is evident to prevent pollution of online discourse and thwart political information operations \cite{Lazer}. Analyzing the state-sponsor trolling linked to the Russian troll farm Internet Research Agency (IRA), a study found that trolls create a small portion of original trolling content (e.g. posts, hashtags, memes, etc.) and heavily engage in retweeting around a certain point in time of interest (e.g. the Brexit referendum) \cite{Llewellyn}. A detailed investigation into the trolling activity around the 2016 U.S. elections reveals different state-sponsored trolling with varying tactics: IRA trolls were pro-Trump while Iranian trolls were anti-Trump \cite{Zannettou}. Authors in \cite{Jane} trained a classifier on a Twitter-released IRA trolling dataset and tested it on sample of accounts engaging with prominent journalists and were able to distinguish between a troll and a non-troll with a precision of 78.5\% and an AUC of 98.9\%, under cross-validation. Another trolling detection algorithm analyzing the writing style of the IRA Twitter trolls looked into the emotional, morality, and sentiment changes showing a 0.94 F1 score \cite{Ghanem}.  

The benefits of using Twitter-released trolling narratives as the training dataset -- IRA or other state-sponsor trolls -- are obvious because they include accounts, tweets, hashtags, and links confirmed by Twitter as part of a political information operations campaign after an investigation into violating their terms of use \cite{Roth}. This approach is somewhat limited, however, because this method of automated detection aims to detect \textit{trending} trolling narratives based on their \textit{past} behaviour. State-sponsored troll farms are engineered with the objective of persistence in their information operations, and thus, adapt in response to detection and development of new polarized topics. For example, an automated detection tool trained on the Twitter datasets will likely misclassify most of the trolling narratives proliferated during the COVID-19 pandemic because the tweets, hashtags, and tropes revolve around the false information about ``China's responsibility for the pandemic,'' the ``status of COVID-19 treatment and vaccine,'' and the ``origin of the COVID-19 virus'' as shown in \cite{nspw2020}. To account for this limitation, we identified potentially polarizing events in the context of the 2020 U.S election cycle and tracked trolling narratives on Twitter in real-time to observe how these narratives constantly evolve. For this, we continuously updated our dataset as the 2020 US campaigning and election season unfolded, which allowed our mechanism, called \textit{TrollHunter2020}, to capture the \textit{trending} emerging behaviour as it unfolded on Twitter. 

\subsection{2020 U.S. Elections}
The 2020 U.S. election cycle was characterized by an unprecedented division between the political parties during exceptionally trying times of the COVID-19 pandemic \cite{Dimock}. With many people at home, the public discourse around the 2020 U.S. election cycle naturally took place on social media platforms - with Twitter of special interest serving as a \textit{de facto} press center for the incumbent president Donald Trump. With the previous evidence of information operations on Twitter, and the high stakes of the 2020 U.S. election cycle, researchers began collecting corresponding Twitter data \cite{Chen}, available for identifying trends, tropes, bot accounts, and potential information operations. In their work, Ferrara et al. accumulated a massive dataset of tweets to analyse bot/human activity and trolling narratives spread by banned accounts on Twitter \cite{Ferrara}. Their findings indicated highly partisan behavior in retweeting among bots and humans, demonstrating that the political discourse on Twitter was self-reinforcing. Using Twitter’s un-hashed Banned User dataset, Ferrara et al. found information operations activity from state-sponsored trolls interacting both with left-leaning and right-leaning users. The trolling, or what they call ``distorted'' narratives that emerged from their dataset were, expectedly, centered around the QAnon, “-gate”, and COVID-related conspiracy theories. 

In our study, we are also looking for trolling or distorted narratives, but with a focus on how they emerge in real-time, populating during ongoing polarizing events, instead of their association with known conspiracy theories or known/suspected trolling accounts. Our \textit{TrollHunter2020} is focused on real-time monitoring of early trends on Twitter manifested as hashtags, topics, or themes; it requires a modest amount of data as an input of a correspondence analysis, rather then a typical data classification tool. \textit{TrollHunter2020} outputs the relationship between the top nouns and verbs trolls most likely will use to construct the tolling narratives that later could be identified by the solution proposed by Ferrara et al. or the other mechanisms for automatic trolling detection. In a way, \textit{TrollHunter2020} serves as a real-time early detection system of \textit{present, emerging} trolling narratives that can help improve later detection of associated accounts, state-sponsorship, or other details about a particular information operation. To use \textit{TrollHunter2020} one doesn't need to wait for an account to be banned or flagged by Twitter, but instead, can ``tune-in'' right during an event of interest (e.g. a debate) and retrieve a high-level overview of ``how might trolls distort the narrative around this event.'' 

Trolling has also evolved, and strategies for identifying trolling must evolve as well. Unlike in the past, trolling is no longer just an effort by state-sponsored actors to change perceptions in a target country as evidenced in the Twitter-released sets \cite{Twitter, Im}, but it has become a strategy used by people in positions of power to misguide possible voters while using their official platforms \cite{Fuchs}. This means that a trolling narrative becomes a vital component in a strategy for public relations employed by politicians. In other words, what initially was deemed as ``computational propaganda'' has morphed into an actual ``political propaganda online'' \cite{Woolley}. By propagating alternative narratives from positions of power, those narratives become more legitimate, and in turn, gives cadence to regular Twitter users to become active and participate in the development of an alternative narrative, e.g. amplify the political propaganda online, which resembles an organic discourse compared to the trolling narratives manufactured by the state-sponsored troll farms \cite{Wilson}. The seemingly concerted distortion of a narrative, in this form, could naturally evade detection since it does not come from suspected accounts, does not necessarily use trolling hashtags, nor does it depend on linking content from alternative media. Therefore, we concentrated our lookout for emerging trolling narratives on the textual content of the tweets and it's potency to be developed further with the organic Twitter discourse.


\section{TrollHunter2020: Design}
\subsection{Context Challenges}
In building \textit{TrollHunter2020}, we set out to leverage the sophisticated analysis characteristic for the troll hunting models presented above, however, our goal was to achieve an acceptable performance without depending on massive datasets and copious amounts of historical data. The design choice of real-time troll narrative hunting on Twitter, thus, presents a series of unique needs for applying data mining. First, we needed to collect the data in real-time, quickly, as it happens, in order to capture the \textit{trolling opportunity} presented by an event or series of events during the elections -- for example the 2020 vice presidential debate -- before the trolls are tagged for misinformation \cite{Roth} or their accounts banned \cite{Twitter}. Second, we needed to operate on small datasets that can be quickly accumulated while the event is trending on Twitter instead of waiting for Twitter to retroactively investigate accounts which it suspects might be furthering election interference goals \cite{Twitter}. Third, election events like candidate debates usually generate multiple narratives so we needed to be able to distinguish between each narrative, even with a small amount of data. That required a careful choice of the number of tweets so we could start working with in order to achieve meaningful troll hunting for distinguishing narratives associated with a given event. In the case of \textit{TrollHunter2020}, the dataset size can be as few as several dozen to few hundred tweets. 

Fourth, before an event, it was sometimes unclear what we would be looking for as a \textit{trolling narrative}. For example, during the 2020 vice presidential debate, a fly landed on Vice President Pence's head. There is no way for a troll or a troll detection mechanism to prepare \textit{apriori} for that, but \textit{TrollHunter2020} did need to decipher if that would become a dominant trend in the data. This entails use of unsupervised techniques in order to capture the \textit{present trolling narratives} organically emerging from the event instead of using predefined trolling labels. Fifth, the troll hunting analysis needed to run and present meaningful results quickly. That means that the \textit{TrollHunter2020} must (a) avoid experimentation with hyper-parameters since that takes time and (b) the troll hunt analysis must be computationally efficient. This also required a careful synchronisation between the ongoing event, the Twitter API rate, and the corresponding Twitter activity; if we had tried to analyze a specific event that has just started, e.g. the 2020 vice presidential debate, people might not have tweeted enough content yet for us to have gained any meaningful insight, but there potentially could be a burst of tweets or hashtags as the event progressed, e.g. after a heated disagreement on the mail-in voting, that we could not risk missing.  

\subsection{Data Preparation}
\textit{TrollHunter2020} uses the \verb!Tweepy! python library to accumulate tweets corresponding to the specific search terms and hashtags listed in Section 3.5 below for each event of interest. This created datasets where each sample represented one tweet. We used the Twitter search API, so every 3 minutes \textit{TrollHunter2020} could search for tweets, iterating through pre-defined lists of search terms. \textit{TrollHunter2020} then uses the \verb!spacyr! \cite{spacyr} R library to parse the parts of speech from each individual word in each tweet. The \verb!spacyr! broke the tweets into individual words, so each sample became a single word from a single tweet, and for each word parsed the part of speech, along with its lemma (the lemma is a word stem that combines different forms of the same word into a single word, so ``lies,'' ``lied,'' and ``lying'' all become ``lie''). \textit{TrollHunter2020} performs additional text cleaning using the \verb!tidytext! R library \cite{tidytext} to remove all stop words using the built-in stop words list. Stop words are words that provide little analysis-relevant information, like ``the,'' or ``an.'' 

Additionally, for each analysis \textit{TrollHunter2020} removes specific words that do not provide useful information. For example, in the context of the 2020 vice presidential debate, the noun ``question'' and the verb ``answer'' both did not provide much useful information. These words were not necessarily intuitive, and this step, for now, requires manual intervention during an event to achieve a finer data representation. Next,\textit{TrollHunter2020} performs a full outer join for the dataframe itself using the \verb!dplyr! \cite{dplyr} R library to create samples of each noun and verb lemma pair featured in each individual tweet. We filtered the dataset to only include the top 10 verbs and nouns. Finally, \textit{TrollHunter2020} creates a contingency table where each value represents how many times each verb lemma (the samples) appeared with each noun lemma (the variables). This yielded a final dataset for the correspondence analysis only with 100 values.


\subsection{Correspondence Analysis}
\textit{TrollHunter2020} performs the correspondence analysis using the R \verb!FactoMineR! library \cite{FactoMineR} to hunt for trolling narratives in real-time on Twitter during the 2020 U.S. election cycle. Correspondence analysis is a technique for visualizing data related to a couple of categorical variables. Correspondence analysis provides ``a window onto the data, allowing researchers easier access to their numerical results and facilitating discussion of the data and possibly generating hypotheses which can be formally tested at a later stage'' \cite{greenacre}. In other words, a correspondence analysis would ``ideally be seen as an extremely useful \textit{supplement} to, rather than a replacement for, the more formal inferential analysis such as \textit{log-linear} or \textit{logistic} models'' \cite{everitt1}. In the context of decision support for real-time trolling narrative identification, however, we determined that a primarily exploratory technique is most appropriate. In the context of our mechanism \textit{TrollHunter2020}, we are trying to quickly get a high-level overview of the trolling patterns as they are emerging in the data corresponding to an ongoing event in the 2020 U.S. election cycle. As such, the high-level overview provided in a correspondence analysis is ideal for \textit{TrollHunter2020} to act fast and capture the trolling narratives before they vanish or morph on Twitter.

A correspondence analysis is a method for comparing two categorical variables by putting the two variables in a contingency table and then taking the $\chi^2$ statistic, which can be put into a matrix $C$ of elements $c_{ij}$ calculated as:
$$
c_{ij} = \frac{n_{ij} - E_{ij}}{\sqrt{E_{ij}}}
$$
where $E_{ij}$ represents the expected value if all f the values in the contingency table are the same.\cite{everitt1}. To get the first two dimensions from this matrix, we use a matrix decomposition: 
$$
C=U\Delta V'
$$
Where $U$ represents the eigenvectors of $CC'$, $V$ contains the eigenvectors of $C'C$, and $\Delta$ represents the diagonals \cite{everitt1}. The values that we are primarily interested in are the first two columns of $V$, which, for our purposes, we plotted to give us an overview of the data.

\subsection{Data Collection}
As previously discussed, we decided not to work on a Twitter dataset with already identified trolling users and trolling tweets  nor to rely on user's reports of twitter trolls/post like in \cite{Fornacciari}. Instead, we utilized the Twitter API to collect tweets related to key events during the 2020 U.S. election cycle that were sure to spark divisive content on Twitter. \textit{TrollHunter2020} allows edits to our constructed set of target words based on the trending keywords/hashtags on Twitter as well as tweets that mentioned either of the two presidential candidates: ``@realDonaldTrump'' or ``@JoeBiden''. The creation of the dataset for any given event is contextually dependent, therefore, \textit{TrollHunter2020} develops it in real-time so the dataset can be as holistic and representative as possible. For example, during the vice presidential debate on October 7, 2020, when a fly landed on Vice President Pence's head, ``fly'' was added to the dataset, since it became quickly clear that this keyword would be used to inject trolled content into the discourse surrounding the debate, regardless of the irrelevance to the political event itself. We selected approximately twenty different terms/hashtags for each event in order to limit our analysis to the \textit{most} topical issues on Twitter during and after significant events took during the 2020 U.S. elections cycle. 


\section{TrollHunter2020: Real-time Detection of Trolling Narratives}
We collected and analyzed a series of tweets from a few different events that offered a fertile ground for trolling during the 2020 U.S. election cycle. We selected the presidential and vice presidential debates as well as Election Week, because we believe these events generated the most discourse on Twitter, and by extension, the most alternative narratives around the candidates and their campaign platforms. 


\subsection{Vice Presidential Debate}
For the initial iteration of data collection, \textit{TrollHunter2020} required that search terms be only one word, so we had to shorten some phrases down to single words to use them to collect tweets. We collected tweets using the following trending hashtags, themes, and the official and personal Twitter handles of Senator Harris and Vice President Pence:

\begin{itemize}
    \item \textbf{\#BidenHarris2020} a hashtag used to associate a tweet with the Biden/Harris campaign.
    \item \textbf{\#DebateNight} a hashtag used to associate a tweet with the vice presidential debate. 
    \item \textbf{\#PenceKnew} refers to Pence knowing about the seriousness of COVID-19 despite Trump's public downplaying of the events related to the pandemic. 
    \item \textbf{\#VPdebate}
    \item \textbf{abortion} was discussed widely on Twitter as a topic because of Amy Coney Barrett's recent Supreme Court nomination. 
    \item \textbf{blm}, short for ``Black Lives Matter,'' was discussed in the context of police brutality during the debate. 
    \item \textbf{brutality} was used to identify tweets that reference ``Police Brutality.''
    \item \textbf{catholic} in the context of the debate could have referred to two things: first, Former Vice President Biden's Catholic faith, as well as the faith of Amy Coney Barrett, the U.S. Supreme Court Justice nominee that President Trump had nominated the preceding week.  
    \item \textbf{china} was discussed extensively through the debate with regards to trade with China, as well as China as the source of COVID-19, and the candidates' relationships with China.
    \item \textbf{deal} for Green New Deal, a piece of proposed legislation by Representative Alexandria Ocasio-Cortez (``AOC'') and Senator Ed Markey that was discussed by the candidates during the debate. 
    \item \textbf{roe} refers to Roe vs. Wade, which is a 1973 Supreme Court hearing that grants women the right to an abortion, which some progressives feared would come under attack with a conservative Supreme Court, which Amy Coney Barrett, Trump's nominee could have secured.
    \item \textbf{soleimani} referring to the discussion of Qamed Solaimani, a Major General of Iran's Revolutionary Guard Corps and Commander of the Quds Force who had been killed in a drone strike ordered by President Trump.
    \item \textbf{speaking} refers to the instances where Vice President Pence interrupted Senator Harris, and Senator Harris said the words ``I'm speaking.''
    \item \textbf{swine} refers to the ``swine flu'' which Vice President Pence used to compare to COVID-19.
\end{itemize}

\begin{figure*}[!h]
  \centering
  \includegraphics[width=0.7\linewidth]{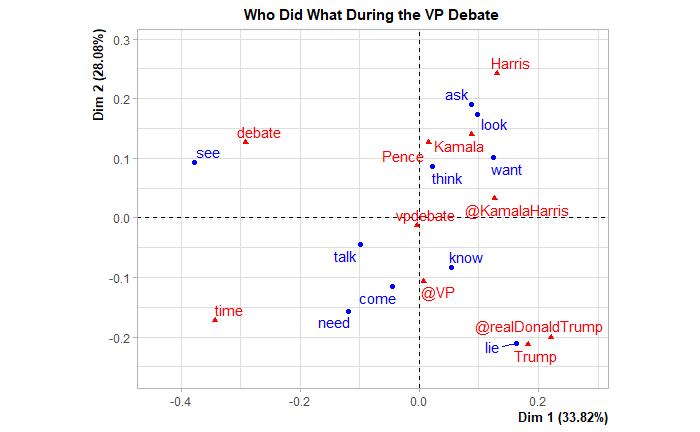}
  \caption{Correspondence Analysis of Top Verbs and Nouns used on Twitter to Describe the Vice Presidential Debate}
\end{figure*}

Figure 1 shows the resulting \textit{TrollHunter2020} correspondence analysis for real-time hunting of trolling narratives on Twitter during the 2020 vice presidential debate. \textit{TrollHunter2020} shows that President Trump (the terms ``@RealDonaldTrump'' and simply ``Trump'') is closely associated with the verb ``lie.'' The association is very strong (both nouns and the verb are far from the origin and the angle between the lines connecting them to the origin is very small, resulting in a higher value for the cosine). This suggests that even though he was not present at the debate, a dominant narrative during the debate is that President Trump lies. This might be an obvious notion, but the utility of \textit{TrollHunter2020} in this case is precisely identifying these relationships and emerging direction of trolling. Trolls are aware that Twitter or anyone tracking their activity will be focused on the salient topics identified above (e.g. \#PenceKnew or \#AOC) and might not look for a hashtag like \#TrumpLiedPeopleDied used towards right-leaning users.


When we look at the overall dimensions, dimension 1 (the x-axis), which accounts for 33.82\% of the inertia of the data, seems to show how political or procedural a noun/verb pair is, with ``time'' and ``debate'' on the left, and the politicians more towards the right. Mentions of ``@VP'' and ``Pence'' are both closer to x=0, while ``Harris,'' ``Kamala'' and ``@KamalaHarris,'' are slightly more to the right, and ``Trump'' and ``@realDonaldTrump'' are both even further to the right. Dimension 2, which accounts for 28.08\% of the inertia in the data, shows Republicans versus Democrats. ``Trump'' is all the way on the bottom, and ``Harris'' and ``Kamala'' all the way on the top. This interpretation of this dimension would indicate that Harris seemed to be more against Trump than against Pence. This interpretation also associates ``time'' with Republicans, and ``debate'' with liberals. Again, \textit{TrollHunter2020} gives and indication on how the trolling narratives might unfold - the strong association between ``look'' and ``Kamala'' or ``Haris'' gives an early warning of the trolling narrative about that she looks upset and nervous leading to a resurgence of hashtags like \#KamalaHarrisDestroyed (initially used during the Democratic candidates' debates in conjunction with the hashtag \#DemDebateSoWhite \cite{Linke}). Another alternative narrative relates to the ``condescending look'' on Senator Harris's face when protesting to Vice President Pence ``I am speaking,'' which is distorted in a narrative presenting her as ``arrogant, rude, smug'' and ultimately a seed for a sexist and misogynistic attacks with hashtags like \#HeelsUpHarris \cite{Tumulty}). 

Finally, it is interesting seeing who is and isn't present. While President Trump was very present in the data, and accounted for two of the identified nouns, Former Vice President Biden did not appear at all in the top nouns, even though one of the terms we used to identify debate-related tweets is the hashtag ``\#BidenHarris2020.'' This means that in the tweets included in the correspondence analysis, Pence may have needed to answer for things that Trump did, but Harris may not have needed to do the same for Biden. This distinction could be vital for early discrimination of bot accounts that act on behalf of a nation-state actor, placing Senator Harris as the main target of the emerging trolling narratives after the vice presidential debate.    


\subsection{Final Presidential Debate} Mirroring the previous iteration, we decided to monitor the second and final presidential debate which took place on October 22. With improvements to \textit{TrollHunter2020}, we were able to search phrases as well as single words/hashtags. Juxtaposing the dataset from the vice presidential debate, the final presidential debate featured more trending phrases/keywords than hashtags which is reflected in the second \textit{TrollHunter2020} dataset. In addition to tracking the Twitter handles of each candidate, we collected tweets involving the following trending hashtags and themes:

\begin{itemize}
    \item \textbf{\#Debates2020} for collecting tweets referencing the 2020 debates
    \item \textbf{\#auditTrump} was trending, referencing the President's alleged tax returns which leaked the weekend previously 
    \item \textbf{@kwelkernbc} to include tweets referencing the debate moderator, Kristen Welker's, Twitter handle 
    \item \textbf{prepaid tax} was a phrase President Trump used during the debate to attest to how much tax he truly paid to the IRS
    \item \textbf{Operation Warp Speed} referencing the Presidential endeavor to quickly develop a vaccine for COVID-19
    \item \textbf{H1N1} to criticize the Obama-Biden administration's handling of the H1N1 virus in comparison to the Trump administration's handling of the COVID-19 pandemic
    \item \textbf{bidencare} was a term Biden coined during the debate to refer to his proposed healthcare plan building off of ``Obamacare''
    \item \textbf{beautiful healthcare} was the phrase used by President Trump to describe his healthcare proposal
    \item \textbf{lowest iq} in the context of the debate referred to a phrase President Trump used to describe immigrants who returned for U.S. court proceedings instead of choosing to live in the U.S. without proper legal authorization
    \item \textbf{who built the cages} was a poignant question asked by President Trump to former Vice President Biden who had criticized the Trump administration immigration detention policy
    \item \textbf{least racist} ``person in this room'' was President Trump's phrase used to defend himself against accusations that his response to the Black Lives Matter movement inflamed racial tensions and emboldened racism 
    \item \textbf{poor boys} was mistakenly referenced by Biden when discussing the white nationalist group ``Proud Boys'' 
    \item \textbf{fracking} to include tweets discussing one of the most controversial discussions of the night, over the position of both candidates on fracking and general environmental policy
    \item \textbf{I know more about wind than you do} was a phrase flaunted by President Trump to critique Biden's knowledge of renewable resources like wind
\end{itemize}

Regarding the large amount of trending quotes from the presidential debate, it was interesting to see that on Twitter, users seemed to be more focused on the precise wording of the presidential candidates versus during the vice presidential debate when users tended to focus more on the topics and issues discussed. In our initial correspondence analysis, shown in Figure 2, in the first dimension (the x-axis), which accounts for 86.09\% of the inertia in the data, the primary trend was a verb/noun pair which was ``build,'' and ``cage.'' Following this early warning from \textit{TrollHunter2020}, ``build,'' and ``cage'' were contextualized in an emerging and prevalent trolling narrative after the presidential debate about ``who build the cages'' targeting President Obama about his decision of building immigration cages, additionally amplified with hashtags like \#ObamaCages and \#DeporterInChief.

\begin{figure*}[!h]
  \centering
  \includegraphics[width=0.7\linewidth]{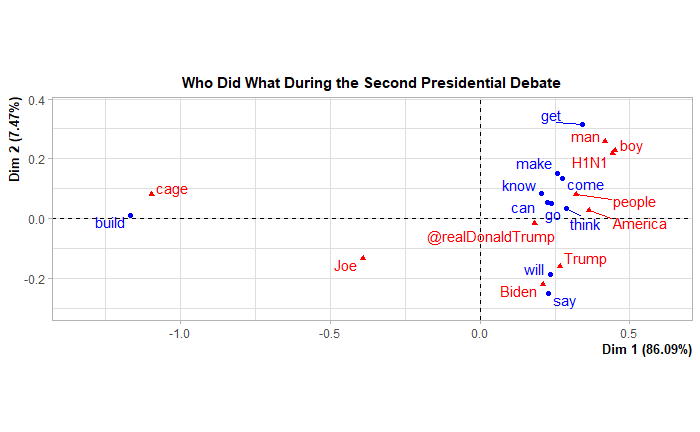}
  \caption{Correspondence Analysis of Top Verbs and Nouns used on Twitter to Describe the Final Presidential Debate}
\end{figure*}%

Using \textit{TrollHunter2020} is an iterative manner, the strong patterns initially identified could be removed to allow for the analysis of additional patterns in the data. To continue analyzing the tweets, We re-ran \textit{TrollHunter2020} without the words ``cage'' and ``build'' with the results shown in Figure 3(a). The first dimension in the correspondence analysis, which accounts for 43.71\% of the inertia in the data, shows for small values the names of the candidates. ``H1N1'' is approximately at Dimension 1 = 0. The term ``come'' and ``man'' could be a reference to Biden's quote ``come on man'' in reference to an interruption by Trump in the first presidential debate. 

The pattern that correlates with higher valued for the first dimension is a little ambiguous, so we elected to remove a few more terms, ``boy,'' ``man'' and ``get'' for a second iteration. As shown in Figure 3(b), the reduced dataset clarified the results further. The words ``Joe'' and ``Biden'' are all the way to the left, ``H1N1'' moves slightly to the left. The removed words make space for ``Obama'' whose name is a lot closer to ``Trump'' and ``@realDonaldTrump'' than it is to ``Joe'' or ``Biden.'' The verb ``take'' and the noun ``fault'' significantly further to the right, indicating that a lot of people were likely trying to attribute blame to Joe Biden about the mass deportations under the Obama administration and his involvement as a Vice President \cite{Devereaux}. 

This \textit{TrollHunter2020} early warning sheds additional light on the emerging trolling narrative targeting Joe Biden's previous White House legacy of more than 3 million immigrants removed from the US \cite{dhs} with hashtags like \#ObamaCabal and \#BidenCrimeFamily. Alternative narratives surrounding the Biden family gained significant traction after President Trump himself tweeted out the hashtag \#BidenCrimeFamiily in a seemingly innocuous attempt to evade Twitter's censorship of the correct hashtag. This points to the usefulness of \textit{TrollHunter2020} to provide concrete warnings of the danger of propagating trolling narratives from positions of power, since it adds additional legitimacy and encourages Twitter followers and supporters to continue to disseminate those views, regardless of the accuracy of the narrative itself.


\begin{figure*}%
    \centering
    \subfloat[\centering ``cage'' and ``build'' removed]{{\includegraphics[width=0.49\linewidth]{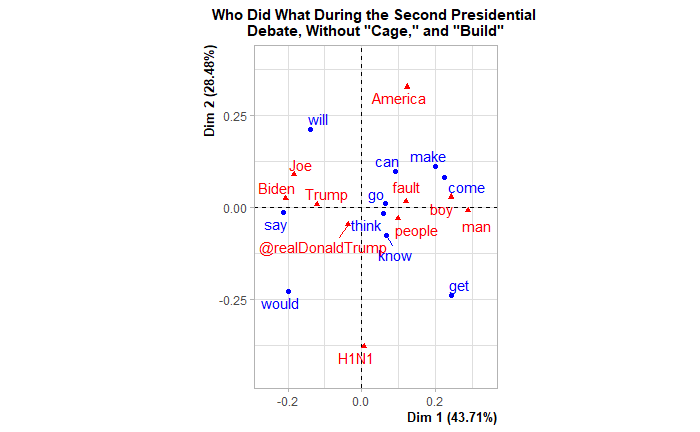}}}%
    \hspace{0.1em}
    \subfloat[\centering ``boy,'' ``man' and ``get'' removed]{{\includegraphics[width=0.49\linewidth]{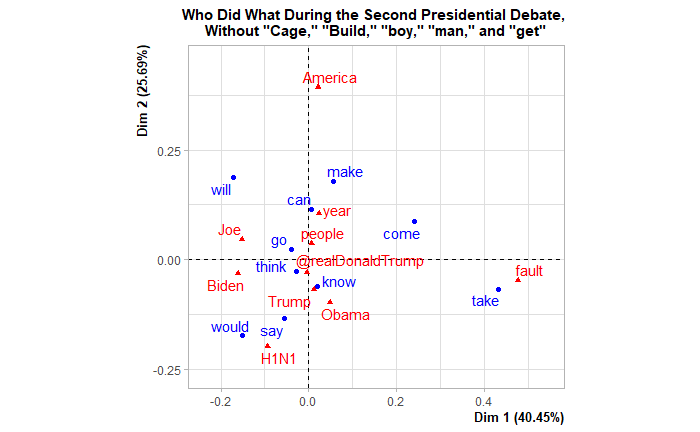} }}%
    \caption{Iterative Correspondence Analysis of Top Verbs and Nouns used on Twitter to Describe the Final Presidential Debate}%
    \label{fig:3}%
\end{figure*}





\subsection{Election Week} Due to the influx of mail-in ballots, unlike previous election cycles, it was expected that a winner of the 2020 U.S. Presidential Election would not be called on Election Night, scheduled for Tuesday, November 3, 2020 \cite{Goldmacher}. Therefore, for the purposes of this event, we collected data about ``Election Night'' throughout the week, leading up to the call by Associated Press (and others) on Saturday, November 7, 2020 that declared former Vice President Biden the projected winner of the 2020 Presidential Election and therefore the President-Elect \cite{AP}. Although the election does not officially end when media outlets project the winner of the race, for the purposes of analyzing public discourse, the projection by news outlets serves as a critical shifting point in conversation regarding the outcome, and therefore, allowed us a break point on when to stop data mining.

\begin{itemize}
    \item \textbf{\#AmericaDecides2020} was the generic hashtag used to refer to everything election-related on November 3, 2020
    \item \textbf{\#ElectionNight} was another hashtag used to reference election-related content on the day of the elections
    \item \textbf{\#TrumpMeltdown} originated in reference to a speech by President Trump on the night of Thursday, November 5 that some perceived as a disheveled, tired, disjointed appearance and message from the president
    \item \textbf{\#StopTheSteal} was a hashtag coined by Trump supporters fearful that the election was stolen by Joe Biden and the Democratic party
    \item \textbf{\#StopTheCount} went viral in the days after Election Day, when mail ballots were still being counted, urging election workers to stop counting ballots
    \item \textbf{\#CountEveryVote} also went viral in the days after Election Day, directly contradicting the previous hashtag, urging election workers to continue to count ballots
    \item \textbf{\#IHaveWonPennsylvania} was a claim falsely made by President Trump before the state of Pennsylvania had concluded its vote-tallying 
    \item \textbf{@realDonaldTrump} to track the incumbent President's personal Twitter handle
    \item \textbf{@JoeBiden} to track the Democratic challenger's personal Twitter handle
    \item \textbf{vote by mail} became a main subject of debate during the election by President Trump who made allegations that it was insecure and contributed to election fraud
    \item \textbf{mail ballot fraud} to track tweets that pushed unsubstantiated narratives of vote-by-mail fraud
    \item \textbf{poll watch}-ing was something that was pushed by President Trump in order to provide additional scrutiny in polling places but it became controversial for fear of poll watchers intimidating voters
    \item \textbf{victory} included to track any premature declarations of victory by either candidate or either candidate's supporters before the race could be accurately projected
    \item \textbf{count} was debated on Twitter throughout Election Week, so by using the word ``count'' in our dataset, \textit{TrollHunter2020} could identify content related to ``stop the count'' as well as ``count all votes''
    \item \textbf{rigged election} to track tweets spreading divisive information that the election was ``rigged'' or illegitimate
    \item \textbf{steal election} was utilized to identify tweets accusing any party of perpetrating the ``rigging'' of the election
    \item \textbf{magically} was the term used by President Trump to describe the vote tallying on Election Night that ultimately swung the vote in key swing states toward Joe Biden
\end{itemize}

\begin{figure*}
  \centering
  \includegraphics[width=0.7\linewidth]{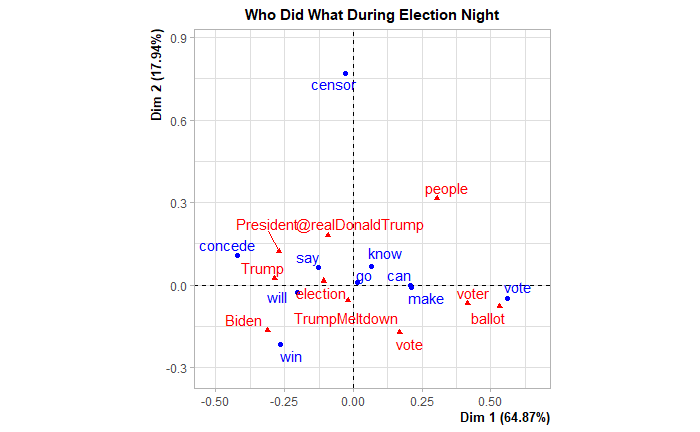}
  \caption{Correspondence Analysis of Top Verbs and Nouns used on Twitter to Describe Election Night}
\end{figure*}

In the Election Night correspondence analysis, The first dimension, which accounts for 64.87\% of the inertia in the data, shows slightly different trends for nouns and verbs. Lower values for nouns in the first dimension show specific politicians, like ``Trump,'' or ``Biden,'' with specific election procedural nouns appearing as higher values, like ``voter,'' and ``ballot.'' Interestingly, ``Biden'' is more likely to be associated with ``ballot'' which feeds to the distorted narrative of ballot harvesting in Texas orchestrated by a high-level staff member on behalf of Joe Biden \cite{Semple}. For verbs, the lower values show terms like ``concede,'' and ``win'' that indicate election results, while more active verbs like ``can,'' ``make,'' and ``vote'' appearing as higher values in the first dimension. In addition, the more obvious the election outcome became, the verb ``concede'' was getting closer to the nouns ``Trump,'' ``President,'' and ``@realDonalTrump.'' This is a confirmation of the culminating tension on the election results with Trump's refusal to concede (e.g. \#NeverConcede, \#StopTheSteal, \#BidenCheated2020, etc) and the apparent win of Joe Biden (e.g. \#BidenWon, \#TrumpConcede, etc).   

The second dimension which accounts for 17.94\% of the inertia in the data shows words more heavily associated with Republicans as higher values. These include nouns like  ``@RealDonaldTrump,'' as well as verbs like ``censor.'' Lower values in the second dimension indicate terms more heavily associated with Democrats, such as ``Biden,'' ``vote,'' and ``TrumpMeltdown,'' which refers to the hashtag ``\#TrumpMeltdown.'' The verb ``censor'' is most closely related to the noun ``@realDonaldTrump,'' which could be a reference to Twitter's decision to label many of Trump's tweets as being potentially misleading \cite{Twitter}. The verb ``will'' is about equally close to both of the candidates, suggesting that in the dataset people were speculating about what both candidates were doing.

\section{Ethical Considerations of Using TrollHunter2020}
We believe that when automated reasoning mechanisms take on cognitive work with social dimensions -- cognitive tasks previously performed by humans -- the mechanisms inherit the social requirements \cite{Frankish}. Social obligations help to highlight the importance of this technique and the \textit{transparency} of the system. This is important because: 1) it builds trust in the system by providing a simple way for users and the wider society to understand what the mechanism is doing and why; and 2) exposes the mechanism’s processes for independent revision \cite{Bryson}. In our case, the main function of the \textit{TrollHunter2020} mechanism was to hunt for trending trolling narratives on Twitter in real-time during the 2020 U.S. election cycle. Like most trolling detection mechanisms, \textit{TrollHunter2020} is far from perfect and incorporates a level of human subjectivity that affects the very definition of what constitutes a ``trending trolling narrative'' in the context of the 2020 U.S. election cycle. Therefore, we developed the \textit{TrollHunter2020} architecture to be generic enough to allow for use of different criteria for alternative or trolling narratives, which in turn enables customization of \textit{TrollHunter2020} for future election cycles or any intense discourse on Twitter. We also ``exposed'' or presented the \textit{TrollHunter2020} to a sufficient degree that allows for independent revision. For interested parties, we are open to share more details about \textit{TrollHunter2020}.

We openly acknowledge that \textit{TrollHunter2020} could be abused in multiple contexts for nefarious purposes with malicious re-purposing of what constitutes an ``alternative, distorted, or trolling narrative.'' For example, in response to popular uprisings in late May 2020, sparked by the killing of George Floyd by police officers in Minneapolis, many activists took to Twitter to spread awareness of the incident, using hashtags like \#BlackLivesMatter, \#BLM, \#iCantBreathe, and \#RestInPower \cite{Rickford}. If \textit{TrollHunter2020} is used without considering the developing socio-political landscape in this case, it might capture combinations of words and nouns that look like an alternative narrative, but might very well be a legitimate trending Twitter topic about a mass struggle. The narrow use of \textit{TrollHunter2020} might prevent this narrative to achieve the intended goal of social action (in fact, the Black Lives Matter (BLM) movement was born out of the \#BlackLivesMatter hashtag on Twitter), which is the opposite effect of what \textit{TrollHunter2020} was developed to achieve. Additionally, with a tool like \textit{TrollHunter2020}, opposition groups have an opportunity to craft emerging counter-narratives in real-time in public spaces like Twitter by combining opposite matching of the noun-verb or noun-noun relationships. For example, \#AllLivesMatter emerged out of retaliation to the initial BLM narrative and movement. Therefore we strongly recommend the use of \textit{TrollHunter2020} only as an early, high overview troll hunter with a consideration of the contemporary socio-political landscape in deciding narratives that actually pollute the public discourse.

\section{Discussion}
\subsection{Implications of Using TrollHunter2020}
Zannettou et al. showed that alternative narratives flow through multiple social media platforms like Reddit, Twitter, and 4chan \cite{Caulfield}. The introduction of real-time trolling narrative detection, in general, could potentially have an effect of moving trolls to less regulated platforms. A recent example of such a migration from Twitter to Parler, Rumble and Newsmax was witnessed after Twitter actively labeled and removed false information on the platform during the 2020 U.S. elections \cite{Isaac}. An opposite effect is also possible, where trolls or fringe Web communities disseminating trolling narratives could be attracted on Twitter by exploiting the limitations of \textit{TrollHunter2020} to precisely identify the potent or most probable candidate of words for the emerging trolling narratives. Aware of the limitation of the correspondence analysis of \textit{TrollHunter2020}, trolls could come up with alternative tropes or words, or hashtags, for example, ``demagogue'' instead of ``lie'' (for the vice presidential debate), ``blue-pencil'' instead of ``censor'' (for the final presidential debate), and ``triumph'' instead of ``win'' (for the Election Night), that could evade detection or throw off the attention.

\subsection{Inherent Limitations of TrollHunter2020}
\textit{TrollHunter2020} like every automated detection mechanism comes with a set of limitations. Even though we attempted to capture the most popular topics on Twitter surrounding the events we reviewed, they might not represent the complete picture around a given election event, especially around events that are discussed not just on Twitter but on many other social media platforms and forums. The selection of hashtags, themes, and accounts to be involved was rather limited, and one might yield different lists of top words to be used for the correspondence analysis if they employ a different selection criteria. \textit{TrollHunter2020} provides a real-time high overview of emerging trolling narratives and is possible to misidentify or completely miss a combination of words, hashtags, and accounts. \textit{TrollHunter2020} does not identify trolling accounts nor discriminates between types of accounts, e.g. bot versus human accounts. \textit{TrollHunter2020} highlights some of the biggest narratives circulating, and should only be used to help clarify and analyze the overall trends in the data in a decision support role rather than an automated hunter of trolls on Twitter.

\subsection{Future Adaptations of TrollHunter2020}
Juanals and Minel used a very structured analysis to analyze the dissemination of information across different Twitter accounts as an approach to supplement their correspondence analysis \cite{juanals}. It might be interesting to incorporate some of these more structured elements into a future iteration of \textit{TrollHunter2020} to see the outcome of the troll hunt. For example, \textit{TrollHunter2020} could be enhanced to to track the spread of messaging from the President to their cabinet members' Twitter accounts or from any person of interest during and election and their closes collaborators on Twitter. Another possible adaptation of \textit{TrollHunter2020} is to yield a comparative correspondence analysis between two datasets created around a highly publicized event, for example,  a Twitter dataset and a Parler dataset. The benefit of such a comparison is to broaden the scope of detection of early trolling narratives, given that social media becomes more segmented based on ideological preferences and as a result of active trolling detection from Twitter. In this context, it would be interesting to employ \textit{TrollHunter2020} to monitor the ongoing Twitter discourse around the approval, distribution and administration of the COVID-19 vaccines and treatments.     

\section{Related Work}
Interestingly, correspondence analysis as an approach to data analysis was primarily used by French researchers for an extended period of time \cite{everitt1}. This may have impeded our ability to do as robust of a literature review as we may have liked, but based on our review, no one has applied correspondence analysis in the way that we have. Juanals and Minel applied a similar technique to show tweets by different art museums during the French Night of the Museums in 2016 \cite{juanals}. In their work, Juanals and Minel tracked messaging and message dissemination through a pre-defined list of accounts to show how messages on Twitter spread between specific museums and their stakeholders.

Unlike Juanals and Minel's approach, which requires a certain amount of structure and predictability to be present in the events that they analyze, due to the our pre-processing pipeline, and our analysis of correlation between verbs and nouns, our approach is relatively open and can be implemented more quickly in less predictable circumstances. Our approach does not factor in the relative popularity of some accounts over others, nor do we define stakeholders, or which accounts might tweet topics of interest. Several other papers used correspondence analysis based on Tweet metadata or manually derived features. For example,  Maity, Saraf, and Mukherjee presented a strategy for predicting which hashtags would perform better when combined \cite{maity}. They used a correspondence analysis to demonstrate the success of their techniques based on features they had engineered to show how their approach has a higher success than humans in predicting successful compound hashtags \cite{maity}.

\section{Conclusion}
In this paper we present \textit{TrollHunter2020}, a simple approach to assisting in the analysis of real-rime detection on trolling narratives on Twitter during the 2020 U.S. election cycle. Unlike other automated trolling detection mechanisms, \textit{TrollHunter2020} doesn't require a massive amount of data for classification nor is dependent on subjective labeling. \textit{TrollHunter2020} feeds of the trending topics on Twitter around a highly publicized election event, a debate, election day, vote counting, etc. and employs a correspondence analysis to yield the strongest relationships between the most probable combinations of words (nouns and verbs) that will be utilized in trolling following the event of interest. During the vice presidential debate, \textit{TrollHunter2020} detected a follow-up trolling narrative based on the allegedly ``condescending look'' of Senator Harris alluding to her being ``destroyed'' in the debate by Vice President Pence. During the final presidential debate, \textit{TrollHunter2020} pointed to a trolling narrative involving Joe Biden's role as a vice president in the Obama administration in building ``cages for kids and immigrants'' and the removal of a record number of immigrants outside of the U.S. During the election-night-turned-election-week, \textit{TrollHunter2020} pointed to the interplay between the distorted narratives of Trump not conceding and Joe Biden's involvement in election fraud. \textit{TrollHunter2020} comes with limitations and we suggest its use for real-time trolling narrative detection only in consideration with the contemporary socio-political context of any polarized Twitter debate. Notwithstanding this, we believe that \textit{TrollHunter2020} can be used in future analyses to enhance troll hunters' abilities to quickly detect trends in Twitter data with less information to help them make use of their findings more quickly and efficiently.

\bibliographystyle{IEEEtran}
\bibliography{trollhunter2020}

\end{document}